# Revisiting the (110) Surface Structure of TiO$_2$: A Detailed Theoretical Analysis

Scott J. Thompson and Steven P. Lewis
*Department of Physics and Astronomy, Center for Simulational Physics,
University of Georgia, Athens, Georgia 30602-2451*

A detailed reexamination of the (110) surface structure of rutile TiO$_2$ has been carried out using first-principles total-energy methods. This investigation is in response to a recent high-precision LEED-IV measurement revealing certain significant quantitative discrepancies between experiment and previous theoretical calculations. We have been able to resolve these discrepancies and achieve excellent quantitative agreement with experiment by judicious attention to reducing computational approximation errors. Our analysis also reveals that bond lengths converge *much* faster with slab thickness than do relaxed absolute atomic positions, which are the structural parameters typically reported in the literature for this and related systems. The effect this observation has on both the choice of slab models and the way in which surfaces structures should be reported for covalently bonded solids is discussed. Finally, the efficacy of freezing the lowest several atomic layers of TiO$_2$ slab models in their bulk-like positions is examined.

A recent LEED-IV experiment[1] has produced what is arguably the highest precision measurement to date of the structure of the (110) surface of rutile TiO$_2$. Comparison of these data to various published first-principles theoretical investigations of this system[2-8] reveals that, while some predicted atomic positions are in good agreement with the new experiment, others – most noticeably those of the surface oxygen atoms – are not. Indeed, the vertical relaxation of the bridging oxygen atoms was consistently predicted to be *negative* (i.e., inward relaxation relative to bulk termination), but was measured to be *positive* in Ref. [1]. An earlier surface x-ray diffraction study[9] had reported a significant inward relaxation for the bridging oxygen. Thus, the relaxed structure of the surface oxygen atoms has been a source of confusion both experimentally and theoretically.

Given the prominence of TiO$_2$ (110) in applications and basic research,[10] as well as the extensive number of first-principles investigations on systems that build on TiO$_2$ (110) as a starting point, we felt this confusion was significant and merited further inquiry. As a result, we have undertaken an extensive first-principles analysis of the structure of TiO$_2$ (110), focusing on convergence and characterization issues. Our calculations resolve the discrepancy between theory and experiment on the relaxations of the surface oxygen atoms, while maintaining agreement in the other atomic relaxations. Furthermore, we find that much thicker slabs than have been typically used are necessary to achieve convergence not only in surface energy, as found previously,[11] but also in absolute atomic positions. However, if bond lengths are chosen as the "figures of merit" for assessing convergence, then our results show that the thinner slabs used to date in theoretical studies of this system are acceptable. We argue below in favor of switching to reporting bond lengths (and perhaps bond angles) as a more physically relevant description of the surface.

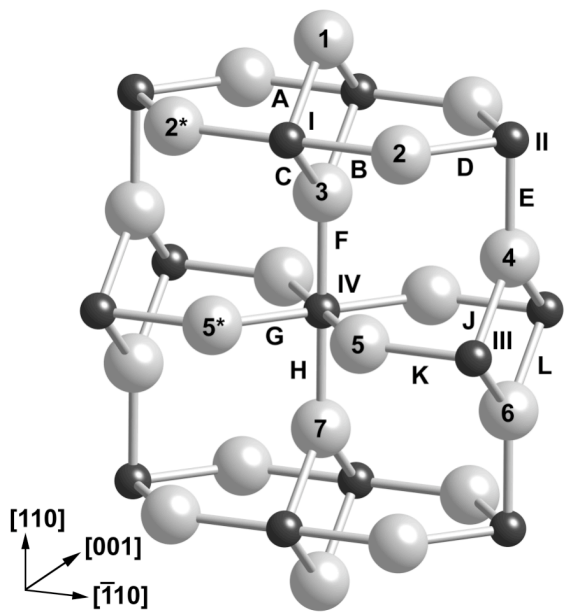

FIG. 1. Model of the TiO$_2$ (110) surface showing 3 trilayers. Atoms in the two trilayers nearest the surface are numbered using Arabic numerals for O atoms (large, light spheres) and Roman numerals for Ti atoms (small, dark spheres). An * denotes an atom paired via symmetry to the atom of the same number. The letters A through L (omitting I) label bonds.

Our *ab-initio* calculations have been carried out within the framework of density functional theory in the generalized gradient approximation using the Vienna *Ab-initio* Simulation Package (VASP).[12-14] Ultrasoft pseudopotentials are used to model core-valence interactions for O and Ti atoms, with the semi-core 3$p$ electrons of Ti included in the valence manifold. Plane waves up to a kinetic-energy cutoff of 396 eV are used for expanding the Kohn-Sham wave functions. Electronic and structural relaxations are carried out using a blocked Davidson-like algorithm and a conjugate-gradient algorithm, respectively, with atomic positions considered relaxed only when all atomic force components are smaller than 10 meV/Å. Other computational specifications in our calculations include: Gaussian broadening, reciprocal space projection, and avoidance of FFT wrap-around errors. Brillouin zone sampling and structural model specifications, such as slab thickness and vacuum spacing, are discussed below.

To set up our surface calculations, we first compute the optimal structural parameters for bulk rutile $TiO_2$. The calculations are found to be well converged with respect to $k$-point sampling using a 4×4×6 Monkhorst-Pack[15] grid in the Brillouin zone. We obtain optimal lattice constants of 4.611 (4.5936) Å and 2.971 (2.9587) Å for $a$ and $c$, respectively, and optimal internal coordinate, $u$, of 0.304 (0.3048). These values compare favorably with experiment[16] (in parentheses) and other theoretical calculations.[3,4,6,7]

In our surface calculations, the system is modeled as a finite slab of stacked $TiO_2$ trilayers periodically reproduced normal to the surface, with 19.6 Å of vacuum separating adjacent slabs. This vacuum spacing, corresponding to the width of ~6 trilayers, is found to be very well converged. Each $TiO_2$ trilayer comprises a central TiO plane with a sparse layer of so-called 'bridging' oxygen atoms symmetrically placed above and below the central plane. Figure 1 shows one unit cell of a three trilayer slab. The number of trilayers in the slab model is an important convergence parameter for the calculations, as discussed below.

Convergence with respect to $k$-point sampling has been studied for 2×2×2, 2×4×2, and 4×8×2 Monkhorst-Pack grids in the Brillouin zone of the 5 trilayer $TiO_2$ (110) slab unit cell. We find the 2×4×2 grid to be the smallest converged $k$-point set, in contrast with previous calculations,[3,4,6,7] which all used a 2×2×2 $k$-point set. This discrepancy partially accounts for the disagreement between the current experiment and the theoretically predicted positions of the surface oxygen atoms. Unless otherwise noted, all slab calculations reported here use the 2×4×2 Monkhorst-Pack grid.

As discussed recently by Bredow, *et al*.,[11] fully converged slab calculations for $TiO_2$ (110) require 11-13 trilayers. We have confirmed this result by computing the surface energy and atomic-position relaxations relative to bulk for slabs containing an odd number of trilayers, up to 15. Our calculation of surface energy vs. slab thickness is shown in Fig. 2, with results of two previous calculations plotted, as well. We find that surface energy is not converged to within the precision of the calculations until 13 trilayers are used.

Our calculations of relaxed atomic positions for slabs containing 5-13 trilayers are summarized in Table 1. Also included are data from two experiments,[1,9] as well as results from one earlier theoretical study[3] on a 7-trilayer slab. Relaxed positions are reported as displacements (in Å) from ideal bulk positions, as measured from the central trilayer of the slab in the theoretical studies and (apparently) from the third trilayer from the surface in the two experimental studies.[1,9] We find that atomic relaxations are converged to within 0.01 Å only for slabs containing 9 or more trilayers. Thinner slabs give significantly poorer convergence.

However, as Table I shows, our results for a 5-trilayer slab agree very well with the recent LEED-IV experiment.[1] In particular, our computed vertical relaxations of the two surface oxygen atoms, O(1) and O(2), are very close to the experimental values. Recall that the largest discrepancies, by far, between the recent experiment[1] and earlier first-principles calculations[2-8] were for these two degrees of freedom. These observations raise two questions that will shape the remainder of this paper: First, why do the present calculations differ from earlier ones, resulting in better overall agreement with the experiment in Ref. [1]? And second, why do calculations of relaxed atomic positions for 5-trilayer slabs agree so closely with experiment even though the convergence for both surface energy and atomic positions has been shown to be very poor for this slab thickness?

The answer to the first question is a matter of inadequately converged computational approximations. As the column labeled 7* in Table I shows, we were able to reproduce the relaxation results of the earlier calculation by Bates et al.,[3,4] to within ±0.03 Å on *every* atomic degree of freedom, by reducing the $k$-point set to 2×2×2, moving the Ti 3$p$ states from the valence manifold to the core, using real-space projection operators, and allowing some degree of FFT wrap-around, as specified by the Normal setting of the PREC input parameter in VASP.[14] Each of these modifications corresponds to worsening the computational approximations. Thus, we see that the closer agreement of the present calculations to the latest experimental results can be attributed to improved computational convergence compared to previous calculations.

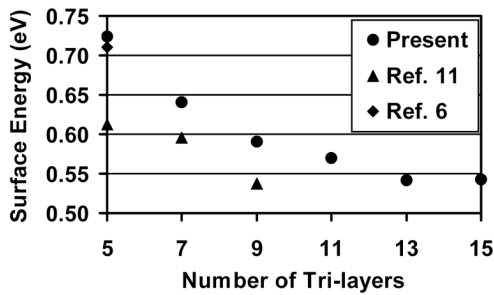

FIG. 2. Surface energy versus the number of trilayers in the slab. Only slabs with odd numbers of trilayers were considered, in order to avoid odd-even oscillations.[11]

| Atom Label | Displacement from bulk positions (Å) | | | | | | | |
|---|---|---|---|---|---|---|---|---|
| | Exp't. (1997)[9] | Calc. (1997)[3] | Exp't. (2005)[1] | Present Calc. (No. of trilayers in slab) | | | | |
| | | | | 5 | 7 | 9 | 11 | 7* |
| Ti (I) | 0.12 ± 0.05 | 0.23 | 0.25 ± 0.03 | 0.29 | 0.33 | 0.42 | 0.43 | 0.20 |
| Ti (II) | -0.16 ± 0.05 | -0.11 | -0.19 ± 0.03 | -0.10 | -0.07 | -0.03 | -0.03 | -0.14 |
| O (1) | -0.27 ± 0.08 | -0.02 | 0.10 ± 0.05 | 0.09 | 0.13 | 0.23 | 0.23 | -0.02 |
| O (2) | 0.05 ± 0.05 | 0.18 | 0.27 ± 0.08 | 0.24 | 0.27 | 0.32 | 0.32 | 0.16 |
| O (2) [$\bar{1}10$] | 0.16 ± 0.08 | 0.05 | 0.17 ± 0.15 | 0.06 | 0.05 | 0.05 | 0.05 | 0.05 |
| O (2*) [$\bar{1}10$] | -0.16 ± 0.08 | -0.05 | -0.17 ± 0.15 | -0.06 | -0.05 | -0.05 | -0.05 | -0.05 |
| O (3) | 0.05 ± 0.08 | 0.03 | 0.06 ± 0.10 | 0.07 | 0.11 | 0.19 | 0.19 | 0.00 |
| Ti (IV) | 0.07 ± 0.04 | 0.12 | 0.14 ± 0.05 | 0.17 | 0.23 | 0.32 | 0.32 | 0.09 |
| Ti (III) | -0.09 ± 0.04 | -0.06 | -0.09 ± 0.07 | -0.04 | -0.03 | 0.00 | 0.00 | -0.09 |
| O (4) | 0.00 ± 0.08 | 0.03 | 0.00 ± 0.08 | 0.06 | 0.09 | 0.13 | 0.13 | 0.00 |
| O (5) | 0.02 ± 0.06 | 0.00 | 0.06 ± 0.12 | 0.06 | 0.09 | 0.15 | 0.15 | -0.02 |
| O (5) [$\bar{1}10$] | -0.07 ± 0.06 | -0.02 | -0.07 ± 0.18 | -0.03 | -0.02 | -0.01 | -0.01 | -0.02 |
| O (5*) [$\bar{1}10$] | 0.07 ± 0.06 | 0.02 | 0.07 ± 0.18 | 0.03 | 0.02 | 0.01 | 0.01 | 0.02 |
| O (6) | -0.09 ± 0.08 | 0.03 | 0.00 ± 0.17 | 0.03 | 0.07 | 0.10 | 0.11 | 0.00 |
| O (7) | -0.12 ± 0.07 | 0.00 | 0.01 ± 0.13 | 0.02 | 0.04 | 0.10 | 0.09 | 0.00 |

Table I. Relaxed absolute atomic positions, given as displacements from bulk-like positions for a hypothetical bulk-terminated TiO$_2$ (110) surface. Results from the present study are shown for several slab thicknesses (denoted by number of trilayers). The column labeled 7* is a calculation of a 7-trilayer slab designed to match the results of Ref. [3] (see text for details). Also shown are the results of two experiments[1,9] and one previous DFT calculation.[3] Atom labels are taken from Fig. 1, and displacements are in the [110] direction, unless otherwise noted. Positive and negative senses are specified by the coordinate axes shown in Fig. 1.

Answering the second question posed above is a bit subtler. The convention for reporting vertical structural relaxations near the surface for this and related systems is to state the displacement of the atoms from the corresponding bulk-like positions *relative to some fiducial plane parallel to the surface*. For slab calculations in which all atomic degrees of freedom are permitted to relax, this reference plane is necessarily the central plane of the slab. For experiments, the data analysis is built around a model in which some plane of atoms is assumed fixed, and the positions, relative to this plane, of all atoms closer to or at the surface are fitted to the data. In most experiments[1,9] on TiO$_2$ (110), the fiducial plane appears to be chosen as the third TiO layer from the surface, with only the outermost two trilayers fitted to the data. These conditions are most similar to the 5-trilayer slab, for which the central plane *is* the third TiO layer from the surface. Thus atomic positions are measured from the same reference plane for the 5-trilayer slab as for the experiments. Had the experimental analysis used the fourth TiO layer as the reference plane, then the most appropriate slab model would be a 7-trilayer slab, and so on. For slabs thicker than 5 trilayers, where atomic positions are measured relative to a deeper reference point than that of the experiments, vertical-relaxation errors of the deeper atoms compound the errors of the atoms at or near the surface.

Nevertheless, even though the experimental model and the 5-trilayer slab share the same reference plane, it is not clear why the agreement between theory and experiment is as good as it is, since the 5-trilayer slab appears to be poorly converged. To understand this, we reformulate the data of Table I to show bond lengths instead of relaxed atomic positions. These results, tabulated in Table II with bond labels as defined in Figure 1, show that bond lengths converge *much* faster with slab thickness than do atomic-position relaxations. Indeed, even for the 5-trilayer slab, most bond lengths are already converged to within 1%. The worst performer is bond *H*, but this is perhaps not surprising, since this bond links the central trilayer to the one above it. In the 7-trilayer slab, even bond H is fairly well converged. The reason atomic-position relaxations converge much more slowly with slab thickness than bond lengths is that atomic positions in covalently bonded solids can be very sensitive to small errors in bond angles, due to trigonometric factors.

| Bond Label | Exp't. (1997)[9] | Calc. (1997)[3] | Exp't. (2005)[1] | Bond Length (Å) Present Calc. (No. of trilayers in slab) | | | | | |
|---|---|---|---|---|---|---|---|---|---|
| | | | | 5 | 7 | 9 | 11 | 13 | 4(2) |
| A | 1.71 ± 0.07 | 1.80 | 1.85 | 1.84 | 1.84 | 1.84 | 1.84 | 1.84 | 1.84 |
| B | 2.15 ± 0.09 | 2.04 | 2.15 | 2.04 | 2.04 | 2.04 | 2.04 | 2.03 | 2.04 |
| C | 1.99 ± 0.09 | 2.09 | 2.08 | 2.11 | 2.11 | 2.12 | 2.12 | 2.12 | 2.11 |
| D | 1.84 ± 0.05 | 1.95 | 1.90 | 1.92 | 1.93 | 1.93 | 1.92 | 1.93 | 1.92 |
| E | 1.84 ± 0.13 | 1.85 | 1.79 | 1.83 | 1.82 | 1.82 | 1.82 | 1.81 | 1.82 |
| F | 1.97 ± 0.12 | 1.90 | 1.90 | 1.89 | 1.86 | 1.86 | 1.86 | 1.85 | 1.87 |
| G | 1.99 ± 0.05 | 1.97 | 2.00 | 1.98 | 1.97 | 1.97 | 1.97 | 1.97 | 1.97 |
| H | 2.18 ± 0.11 | 2.11 | 2.11 | 2.13 | 2.17 | 2.21 | 2.21 | 2.21 | 2.19 |
| J | 2.00 ± 0.08 | 2.02 | 2.01 | 2.02 | 2.04 | 2.05 | 2.05 | 2.05 | 2.04 |
| K | 1.92 ± 0.06 | 1.97 | 1.92 | 1.96 | 1.96 | 1.97 | 1.97 | 1.97 | 1.97 |
| L | 1.94 ± 0.06 | 1.90 | 1.89 | 1.91 | 1.90 | 1.90 | 1.89 | 1.89 | 1.90 |
| Adsorption Energy of Cu (eV) | | | | 1.72 | 1.68 | 1.66 | 1.66 | 1.65 | 1.68 |

Table II. Relaxed bond lengths for atom pairs at and near the TiO$_2$ (110) surface. Results from the present study are shown for several slab thicknesses (denoted by number of trilayers). Also shown are the results of the two experiments[1,9] and one previous DFT calculation.[3] Bond labels are taken from Fig. 1. The slab label 4(2) is defined in the text. The last line gives the binding energy per atom for a monolayer of Cu atoms adsorbed atop the surface bridging O atoms.

We believe that the foregoing results suggest that bond-length convergence is a better criterion for judging slab models than atomic-position relaxations. Bond lengths are more physically relevant than absolute atomic positions, because they do not depend on an arbitrary fiducial reference point. Moreover, the nature of the physical and chemical environment of, say, a surface site is more closely correlated with local structure (*e.g.*, bond lengths and bond angles of nearby atoms) than with absolute atomic positions.

The rapid convergence of bond lengths with slab thickness is very satisfying, as it supports the use of very thin slabs in calculations involving TiO$_2$ (110). This is essential for studying more complicated structures in this system, such as adsorbates or defects, which require supercells parallel to the surface. A common practice for making thin-slab calculations even more efficient is to freeze the lowest *m* atomic layers in their bulk-like positions for an *n*-layer slab. The idea is that the *n*–*m* layers that are free to relax will see a more bulk-like environment below them than if all *n* layers can relax. Two major advantages of this approach are that even thinner slabs may be used, in principle, and that there are significantly fewer degrees of freedom. There are disadvantages, as well: The mirror symmetry of the slab about its central plane is lost, and the ability to compute surface energy becomes less transparent.

To test this approach for TiO$_2$ (110), we have done calculations for various slabs containing 4-7 trilayers, where the bottom 2-4 trilayers are held fixed. In each case, no fewer than 2 trilayers were free to relax. We found that *all* slab combinations studied gave well converged bond lengths. The last column of Table II shows our results for the smallest slab studied, which we designate as 4(2), because it contains 4 trilayers with the bottom 2 held fixed. We feel these results give strong evidence in favor of using a 4(2) slab in future studies of more complicated structures involving TiO$_2$ (110).

As a final test of the performance of slab models for TiO$_2$ (110), we have briefly examined their efficacy in studies of adsorbates. In particular, we have computed the binding energy for a layer of Cu atoms adsorbed atop bridging oxygen atoms. Our results, given in the last row of Table II, show that the Cu adsorption energy is already converged to within 2% for the 7-trilayer slab. Moreover, the 4(2) slab yields the same Cu adsorption energy as the 7-trilayer slab, indicating that the former is suitable for future studies of adsorption on TiO$_2$ (110).

In conclusion, our first-principles calculations of relaxed atomic positions for the TiO$_2$ (110) surface, using a 5-trilayer slab model, are in excellent agreement with the values reported in a recent highly accurate LEED-IV experiment.[1] In contrast with earlier theoretical studies, we have even obtained close agreement for the vertical relaxations of the two surface oxygen atoms, O(1) and O(2). This improved accuracy is attributed to better converged *k*-point sampling, as well as improvements in several other computational approximations. Despite this close agreement with experiment, we have found, in accord with previous authors,[11] that slabs *much thicker* than the 5-trilayer slab are required to achieve converged atomic-position relaxations and surface energies. To resolve these seemingly conflicting outcomes, we have shown that bond lengths converge *much* faster with respect to slab thickness than do absolute atomic positions. Indeed, the 5-trilayer slab is shown to be well converged if bond lengths are used as the figures of merit. We argue in favor of reporting bond lengths (and bond angles) in surface studies of covalently bonded solids, instead of absolute atomic positions, because not only are these quantities more rapidly convergent, they are also more physically relevant. Finally, we have shown that the 4(2) slab model, containing four trilayers with the bottom two held fixed, yields well converged values for bond lengths, as well as adsorption energy for a test adsorbate system.

This research was supported by the NSF under Grant No. DMR-0094199.

[1] R. Lindsay, A. Wander, A. Ernst, B. Montanari, G. Thornton, and N.M. Harrison, Phys. Rev. Lett. **94**, 246102 (2005).
[2] M. Ramamoorthy, D. Vanderbilt, and R.D. King-Smith, Phys. Rev. B, **49**, 16721 (1994).
[3] S.P. Bates, G. Kresse, and M.J. Gillan, Surf. Sci., **385**, 386 (1997).
[4] S.D. Elliott, and S.P. Bates, Surf. Sci., **495**, 211 (2001)
[5] V.Swamy, J.Muscat, J.D. Gale, and N.M. Harrison, Surf. Sci. **504**, 115 (2002).
[6] A. Vijay, G. Mills, and H. Metiu, J. Chem. Phys. **118**, 6536 (2003).
[7] Y. Wang, and G.S. Hwang, Surf. Sci. **542**, 72 (2003).
[8] H. Sano, G. Mizutani, W. Wolf, and R. Podloucky, Phys. Rev. B, **70**, 125411 (2004).
[9] G. Charlton *et al.*, Phys. Rev. Lett. **78**, 495 (1997).
[10] U. Diebold, Appl. Phys. A **76**, 681 (2003).
[11] T. Bredow, L. Giordano, F. Cinquini, and G. Pacchioni, Phys. Rev. B, **70**, 35419 (2004).
[12] G. Kresse, and J. Furthmüller, Comp. Mat. Sci. **6**, 15 (1996).
[13] G. Kresse, and J. Furthmüller, Phys. Rev. B, **54**, 11169 (1996).
[14] G. Kresse, and J. Furthmüller, VASP the Guide, Wie, Austria, 2003.
[15] H. Monkhorst, and J.Pack, Phys. Rev. B, **13**, 5188 (1976).
[16] S.C. Abrahams, and J.L. Bernstien, J. Chem. Phys. **55**, 3206 (1971).